\documentclass[aps,pra,twocolumn,superscriptaddress,showpacs]{revtex4-1} 
\usepackage{graphicx}
\usepackage{amssymb}
\usepackage{amsmath}
\usepackage{color}

\begin{document}
\title{ Anomalies in the specific heat of a free damped particle: \\The role of
the cutoff in the spectral density of the coupling}
\author{Benjamin Spreng}
\author{Gert-Ludwig Ingold}
\email{gert.ingold@physik.uni-augsburg.de}
\affiliation{Institut f\"ur Physik, Universit\"at Augsburg, D-86135 Augsburg}
\author{Ulrich Weiss}
\affiliation{II. Institut f\"ur Theoretische Physik, Universit\"at Stuttgart,
             D-70550 Stuttgart}


\begin{abstract}
The properties of a dissipative system depend on the spectral density of the
coupling to the environment. Mostly, the dependence on the low-frequency
behavior is in the focus of interest. However, in order to avoid divergencies,
it is also necessary to suppress the spectral density of the coupling at high
frequencies. Interestingly, the very existence of this cutoff may lead to
a mass renormalization which can have drastic consequences for the thermodynamic
properties of the dissipative system. Here, we explore the role which the cutoff
in the spectral density of the coupling plays for a free damped particle
and we compare the effect of an algebraic cutoff with that of a sharp cutoff.
\end{abstract}

\maketitle

\section{Introduction}

In the study of dissipative systems the case of strictly Ohmic damping plays a
prominent role because it implies memoryless damping. However, this model is
the result of an idealization which assumes that the density of the
environmental modes weighted by the coupling strength increases proportional to
the frequency even for arbitrarily large frequencies. In realistic scenarios,
this will not be the case and, on a more formal level, it can give rise to
divergencies. Therefore, one is usually obliged to introduce a high-frequency
cutoff in the spectral density of the coupling. Often one can assume that the
cutoff represents the largest frequency scale in the problem and that the
resulting memory time of the damping is shorter than any time scale of
interest.

While the cutoff in the spectral density of the coupling usually only leads to
quantitative changes of relatively little physical interest, occasionally the
presence of a cutoff can make a qualitative difference. Here, we will consider
such a situation. The thermodynamic properties of a free damped particle at low
temperatures depend significantly on an environment-induced renormalization of
the particle's mass. In particular, there exists a regime, where the mass
renormalization can be negative and thus reduces the mass of the particle. When
the renormalized mass becomes negative, one observes anomalies like a negative
specific heat \cite{Haenggi08}. Such anomalies in the specific heat and in the
entropy are of interest in various contexts
\cite{Florens04,Ingold09b,Zitko09,Campisi09,Campisi10,Sulaiman10,Merker12}.

The mass renormalization alluded to here is due to the suppression of the
density of high-frequency environmental modes and thus is a direct consequence
of the very existence of a cutoff in the spectral density of the coupling.
 
We will consider in this paper the specific heat of a free Brownian particle
subject to a linear environment where the spectral density of the coupling
at low frequencies follows a general power law. At high frequencies we allow
either for an algebraic cutoff which in the simplest case will take the form of
a Drude cutoff or for a sharp cutoff where no environmental modes are assumed
to be present above a certain cutoff frequency. It will become clear that these
two cutoff functions can lead to quantitatively quite different results. In
particular, we will find that for a sharp cutoff the appearance of a negative
specific heat cannot be inferred from the leading low-temperature behavior.

Before embarking on our study, we need to say a few words about the meaning of
a negative specific heat which usually should provoke worries about
thermodynamic instability. However, here we are referring to the specific heat
of a system degree of freedom coupled to an environment. It turns out that
beyond weak coupling, the specific heat of a dissipative quantum system is not
uniquely defined \cite{Haenggi06}. Here, we will base our considerations on the
reduced partition function
\begin{equation}
\label{eq:reduced_partition_function}
\mathcal{Z} = \frac{\mathcal{Z}_\text{S+B}}{\mathcal{Z}_\text{B}}
\end{equation}
where $\mathcal{Z}_\text{S+B}$ is the partition function of the coupled ensemble
of system and bath while $\mathcal{Z}_\text{B}$ refers to the partition function
of the bath alone. In the absence of any coupling between system and bath, the
reduced partition function clearly agrees with the partition function of the
system alone.

We can now employ standard relations of thermodynamics to obtain any
thermodynamic quantity from the reduced partition function
(\ref{eq:reduced_partition_function}). The resulting quantities have a clear
physical significance as the difference between the quantity of system and bath
and the same quantity determined for the bath alone. For the specific heat,
we have \cite{Ingold09}
\begin{equation}\label{eq:specheat}
C = C_\text{S+B}-C_\text{B}\,.
\end{equation}
While each of the terms, $C_\text{S+B}$ and $C_\text{B}$, has to be positive,
their difference can very well become negative and in fact it does under
appropriate circumstances. The physical reason is the suppression of the bath
density of states at low frequencies when the system degree of freedom is
coupled to it \cite{Ingold12}. How strongly the bath density of states will be
suppressed depends directly on the high-frequency cutoff of the spectral
density of the coupling.

We start in Sect.~\ref{sec:damping} by introducing general concepts needed to
describe the damped free particle. In particular, we will introduce the
spectral density of the coupling for the two models with, on the one hand, an
algebraic cutoff and, on the other hand, a sharp cutoff. The corresponding
spectral damping functions will be deduced and their properties will be
presented.  The section closes with a discussion of the mass renormalization
associated with the two models. This quantity will play an important role for
the thermodynamic anomalies in the specific heat. Sect.~\ref{sec:chbds} is
devoted to the change of the bath density of states when the free particle is
coupled to it. Special attention will be paid to the differences between the
two models for the cutoff. In Sect.~\ref{sec:redpartfunc_specheat} we introduce
the reduced partition function of the damped free particle which will
constitute the basis of the calculation of the specific heat. We will
demonstrate how the specific heat can be expressed either in terms of the
spectral damping function or the change in the bath density of states obtained in
Sect.~\ref{sec:chbds}.  Sections~\ref{sec:hightemp} and \ref{sec:lowtemp} are
devoted to the behavior of the specific heat at high and low temperatures,
respectively. We close by presenting our conclusions in
Sect.~\ref{sec:conclusions}.

\section{Free Brownian particle subject to general linear damping}
\label{sec:damping}

\subsection{Spectral density of the coupling}

We assume that the free Brownian particle is moving in one spatial dimension
and is subject to a linear but otherwise general damping mechanism. Its
classical or quantum average velocity then obeys the equation of motion
\begin{equation}
\label{eq:equation_of_motion}
\langle\dot v\rangle(t) + \int_{-\infty}^t\text{d}s\, \gamma(t-s)
\langle v\rangle(s) = 0\,.
\end{equation}
Here, $\gamma(t)$ is the damping kernel which in the following will mostly
appear in the form of its Laplace transform, the spectral damping function
$\hat\gamma(z)$. All properties of the damped free particle can be expressed in
terms of the causal velocity response function $\mathcal{R}(t)$ whose Laplace
transform can immediately be read off from (\ref{eq:equation_of_motion}) as
\begin{equation}
\label{eq:velresplapl}
\hat{\mathcal{R}}(z) = \frac{1}{z+\hat\gamma(z)}\,.
\end{equation}
The two terms in the denominator are associated with the inertia term and the
damping term, respectively.

Although in principle it is sufficient to specify the spectral damping function
$\hat\gamma(z)$, it is useful to consider an explicit model leading to 
(\ref{eq:equation_of_motion}). Doing so will allow to more systematically
define the damping mechanism and to give a more physical interpretation of the
results. A free damped particle subject to linear damping can always be modelled
by a Hamiltonian in which the particle described by its position $Q$ and momentum
$P$ is coupled bilinearly to a set of harmonic oscillators with masses $m_n$
and frequencies $\omega_n$ described by their positions $q_n$ and momenta
$p_n$ \cite{Weiss12},
\begin{equation}
\label{eq:hamiltonian}
H = \frac{P^2}{2M} + \sum_{n=1}^{\infty}\left[\frac{p_n^2}{2m_n}
    +\frac{m_n\omega_n^2}{2}\left(q_n-Q\right)^2\right]\,.
\end{equation}     
In general, the coupling of a system to an environment will lead to a potential
renormalization for which we have accounted here by choosing a translationally
invariant Hamiltonian. Although the masses $m_n$ and frequencies $\omega_n$ give
large freedom to choose a Hamiltonian, it turns out that the only quantity of
relevance for the properties of the free damped particle is the spectral density
of the coupling \cite{Hakim85,Grabert88}
\begin{equation}
\label{eq:spectralDensityOfBathOscillators}
J(\omega) = \frac{\pi}{2}\sum_{n=1}^{\infty}m_n\omega_n^3\delta(\omega-\omega_n)\,.
\end{equation}
In particular, the spectral damping function can be obtained from it according to
\begin{equation}
\label{eq:gammaHatSpectralDensity}
\hat\gamma(z) = \frac{2}{\pi M}\int_0^{\infty}\mathrm{d}\omega \,
                \frac{J(\omega)}{\omega}\frac{z}{\omega^2+z^2}\,.
\end{equation}
A property of the bath which will be of relevance later on is its total mass
\begin{equation}
\label{eq:m_bath}
M_\text{bath} \equiv \sum_{n=1}^{\infty}m_n =
\frac{2}{\pi}\int_0^\infty\text{d}\omega\, \frac{J(\omega)}{\omega^3}\,.
\end{equation}

In the following, we will specify the damping mechanism through its spectral
density. We assume the spectral distribution of bath oscillators to be
continuous and to follow a power law at low frequencies,
\begin{equation}\label{eq:specdenspowerlaw}
J_0(\omega) = M \gamma \omega \left(\frac{\omega}{\omega_\text{c}}\right)^{s-1} \, .
\end{equation}
The exponent $s$ thus specifies the low-frequency behavior. The regimes
$s<1$ and $s>1$ correspond to sub-Ohmic and super-Ohmic damping, respectively,
and $s=1$ is the special case of Ohmic damping. 

To obviate divergences of spectral integrals for observables, the actual
spectral density must fall off sufficiently strong in the limit $\omega\to
\infty$. This may be taken into account by equipping $J_0(\omega)$ with a
cutoff function $f(\omega/\omega_\text{c})$ which approaches unity in the limit
$\omega/\omega_\text{c}\to 0$, and drops to zero sufficiently fast as
$\omega/\omega_\text{c}$ goes to infinity.  Hence we put
\begin{equation}\label{eq:specdens}
J(\omega) = J_0(\omega) f(\omega/\omega_\text{c})   \, .
\end{equation}
Without restriction of generality, the reference frequency $\omega_\text{c}$ in
$J_0(\omega)$ is identified with the cutoff frequency in
$f(\omega/\omega_\text{c})$. In the sequel, we  consider both a smooth and a
sharp cutoff of the spectral bath coupling at high frequencies. For the sake of
simplicity, we use from now on units where $\omega_\text{c} =k_{\mathrm B}
=\hbar=1$.

\subsection{Spectral density with algebraic cutoff (model I)}

To be specific, we choose for the case of a smooth cutoff the algebraic
function $f(x) = 1/(1+x^2)^p$, yielding
\begin{equation}
\label{eq:powerlawSpectralDensity}
J_{\rm ac}(\omega) =  \frac{ J_0(\omega)}{[1+(\omega/\omega_{\rm c})^2]^p}\,.
\end{equation}
We will use the subscript `ac' to indicate that a quantity is taken for
the algebraic cutoff. Occasionally, we omit the subscript when the context
permits it.

The frequency integral (\ref{eq:gammaHatSpectralDensity}) with
(\ref{eq:powerlawSpectralDensity}) is convergent for $s$ in the range
\begin{equation}
\label{eq:range_s_p}
0 < s < 2p+2
\end{equation}
and can be expressed in terms of hypergeometric functions. The resulting
expression with a convergent hypergeometric series in the regime $|z|>1$ is
\begin{equation}
\label{eq:gamhatzlarge}
\begin{aligned}
\hat\gamma(z) &= \frac{\gamma}{\sin(\frac{\pi s}{2})} \frac{z^{s-1}}{(1-z^2)^p} \\
&\;\; +\frac{\gamma}{\pi z}{B(p-\tfrac{s}{2},\tfrac{s}{2})
\,\,{}_2F_1(1,\tfrac{s}{2};1-p+\tfrac{s}{2},z^{-2})}\,.
\end{aligned}
\end{equation}
Here, ${}_2F_1$ denotes the hypergeometric function and the function $B(x,y)$
is Euler's  beta function \cite{NISTHandbook10}.

By use of a linear transformation, the second term in
eq.~(\ref{eq:gamhatzlarge}) can be rewritten as a hypergeometric series which
is convergent in the regime $|z|<1$,
\begin{equation} \label{eq:gamhatzsmall}
\begin{aligned}
\hat\gamma(z) &= \frac{\gamma}{(1-z^2)^p} \Big[\frac{z^{s-1}}{\sin(\frac{\pi s}{2})} 
+\frac{2 p }{\pi }\frac{B(p+1-\frac{s}{2},\frac{s}{2})}{s-2}  \\
&\qquad \qquad \times z\, {}_2F_1(1-p,1-\tfrac{s}{2};2-\tfrac{s}{2}; z^2)\Big]\,.
\end{aligned}
\end{equation}
For integer $p$, the $_2F_1$ function in (\ref{eq:gamhatzsmall}) is a terminating
hypergeometric series
\begin{equation}
\label{eq:gamseriesterminating}
\begin{aligned}
\hat\gamma(z) =&\frac{\gamma}{(1-z^2)^p}\bigg[\frac{z^{s-1}}{\sin(\frac{\pi s}{2})}\\
  &\!\! + \frac{1}{\pi}\sum_{n=0}^{p-1} \frac{(-1)^{n-1}}{n+1-\frac{s}{2}}
       \frac{B(\frac{s}{2}, p+1-\frac{s}{2})}{B(n+1, p-n)}
       z^{2n+1} \bigg]\,.
\end{aligned}
\end{equation}
The forms (\ref{eq:gamhatzlarge}) and (\ref{eq:gamhatzsmall}) or
(\ref{eq:gamseriesterminating})  allow us to easily read off the behaviors of
$\hat\gamma(z)$ for large and small arguments. These will be needed in
Sects.~\ref{sec:hightemp} and \ref{sec:lowtemp} to determine the specific heat
at high and low temperatures, respectively.

Under the tighter constraint 
\begin{equation}
\label{eq:range_s_p_restricted}
0 < s < 2p\,,
\end{equation}
the integral $\int_0^\infty {\rm d}\omega \, J(\omega)/\omega $ is consistently
ultraviolet-convergent. As a result, the leading contribution to
$\hat\gamma(z)$ at high frequencies, $z\gg 1$, is proportional to $1/z$,
\begin{equation}\label{eq:gamht}
\hat\gamma(z) = \frac{2}{\pi M} \int_0^\infty\text{d}\omega\frac{J(\omega)}{\omega}
\frac{1}{z} = \frac{a}{z} \, ,
\end{equation}
where for model I
\begin{equation}\label{eq:acoef1}
a_{\rm ac}(s,p) = \gamma \frac{B(p-\frac{s}{2},\frac{s}{2})}{\pi}\,.
\end{equation}
The $1/z$-term  of $\hat\gamma(z)$ determines the high-temperature behavior of
the damped particle, as we shall see in Sect.~\ref{sec:hightemp}.

On the other hand, the low-temperature properties of the free damped particle
are determined by the low-frequency characteristics of $\hat\gamma(z)$. We
obtain for $|z|\ll 1$
\begin{equation}  \label{eq:gamseriesac}
\hat\gamma_\text{ac}(z) = \frac{\gamma z^{s-1}}{\sin(\frac{\pi s}{2})} (1+p z^2)
+\mu_\text{ac}z  + \lambda_\text{ac}z^3 \, ,
\end{equation}
where the orders $z^{s+3}$, $z^5$, and higher are disregarded. We have
\begin{equation}
\label{eq:massrenorm1}
\mu_{\rm ac}(s,p) =  \frac{2\gamma}{\pi } \,
    \frac{ p \, B(\frac{s}{2}, 1+p-\frac{s}{2}) }{s-2}\, ,  
\end{equation}
and
\begin{equation} \label{eq:coeff3ac}
\begin{aligned}
\lambda_{\rm ac}(s,p) &= -\mu_{\rm ac}(s-2,p)   \\
&= \frac{2\gamma}{\pi} \frac{p B(\frac{s}{2}-1,2+p-\frac{s}{2})}{4-s} \, .
\end{aligned}
\end{equation}

The first term in (\ref{eq:gamseriesac}), in which we have included the leading
cutoff dependence,   describes  frequency-dependent damping.  The second term
adds to the inertial term $z$ in the denominator of $\hat{\mathcal R}(z)$. Its
prefactor $\mu =\Delta M/M$ can therefore be interpreted as an effective change
$\Delta M$ of the particle's mass relative to the bare mass $M$ due to the
coupling to the environment. This mass renormalization term will be discussed
in more detail in Sect.~\ref{sec:massrenormalization}. Finally, the last term
becomes relevant for low temperatures in the regime $0<s<2$ at the particular
damping strength for which $\mu=-1$, as we shall see.

In the limit $s\to 2 m$, where $m$ is a positive integer, both the first  term
and the  term of order $z^{2m-1}$ in the square bracket of
(\ref{eq:gamseriesterminating}) become singular. The singularities cancel each
other, however, and a logarithmic term accrues. For the  particular cases
$s=2$ and $s=4$ we have
\begin{equation}
\begin{aligned}
\hat\gamma(z) &= \frac{\gamma}{\pi} \frac{z}{(1-z^2)^p} \big[\psi(1)-\psi(p) - 2\ln(z)  \\
   & \quad +  (p-1) z^2  \, {}_3F_2(1,1,2-p;2,2;z^2) \big] 
\end{aligned}
\end{equation}
and
\begin{equation}
\begin{aligned}
\hat\gamma(z) &= \frac{\gamma}{\pi}  \frac{z}{(1-z^2)^p}  \Big\{ \frac{1}{p-1} \\
& \qquad  + z^2[2\ln(z)+ \psi(p-1)-\psi(1)-1 ]      \\
&\qquad + \left(1-\tfrac{p}{2}\right) z^4\, {}_3F_2(1,1,3-p;2,3;z^2) \Big\}     \,,
\end{aligned}
\end{equation}
respectively, where $\psi(z)$ is the digamma function \cite{NISTHandbook10}.

\subsection{Spectral density with sharp cutoff (model II) }

Concerning our study to which extent thermodynamic properties depend on the
particular form chosen for the cutoff in the spectral coupling, we contrast the
smooth cutoff function discussed in the previous section with a sharp cutoff
function $f(x) = \Theta(1-x)$, where $\Theta(x)$ denotes the Heaviside step
function. Together with the spectral coupling (\ref{eq:specdenspowerlaw}) we 
have
\begin{equation}
\label{eq:SpecDensSc}
J_{\rm sc}(\omega) =  J_0(\omega)\Theta(1 -\omega/\omega_{\rm c})\,,
\end{equation}
where the subscript `sc' indicates the sharp cutoff. From now on we set again
$\omega_\text{c}=1$. The sharp cutoff in the spectral density
(\ref{eq:SpecDensSc}) is in striking contrast to the smooth cutoff in
eq.~(\ref{eq:powerlawSpectralDensity}).

For the spectral density (\ref{eq:SpecDensSc}), the frequency integral in the
expression (\ref{eq:gammaHatSpectralDensity}) for the spectral damping function
$\hat\gamma(z)$ can again be expressed in terms of hypergeometric functions.
In the frequency regime $|z|>1$ one finds
\begin{equation}\label{eq:gamhchf}
\hat\gamma(z) = \frac{2\gamma}{\pi} \frac{
{}_2F_1(1,\tfrac{s}{2}; 1+\tfrac{s}{2}; -z^{-2})}{s z}\, .
\end{equation}
Convergence of the hypergeometric series in the complementary regime  $|z|<1$
is obtained by virtue of a linear transformation of the $ {}_2F_1$ function in
eq.~(\ref{eq:gamhchf}), yielding
\begin{equation}\label{eq:gamsc}
\hat\gamma(z) = \frac{\gamma z^{s-1}}{\sin(\frac{\pi s}{2})}+\frac{2\gamma}{\pi}
z\frac{{}_2F_1(1,1-\tfrac{s}{2}; 2-\tfrac{s}{2}; -z^2)}{s-2}\, .
\end{equation}
In the particular cases $s=1$ and $s=3$, the spectral damping function reads
\begin{equation}
\begin{aligned}
\hat\gamma(z,s=1) &= \gamma - \frac{2\gamma}{\pi}\arctan(z)\,,\\
\hat\gamma(z,s=3) &= \frac{2\gamma}{\pi}\left[z-z^2 \left( \frac{\pi}{2}   
-\arctan(z)\right)\right]\,,
\end{aligned}
\end{equation}
respectively.

The asymptotic high-frequency expression of $\hat\gamma(z)$ may be written in the 
form (\ref{eq:gamht}) with the coefficient
\begin{equation}\label{eq:acoef2}
a_{\rm sc}(s) = \frac{2\gamma}{\pi s} \, .
\end{equation}

Similar to eq.~(\ref{eq:gamseriesac}), the low-frequency expansion may be
expressed as
\begin{equation}
\label{eq:gamseriessc}
\hat\gamma_{\rm sc}(z) = \frac{\gamma z^{s-1}}{\sin(\frac{\pi s}{2})} 
                +\mu_{\rm sc}z  + \lambda_{\rm sc}z^3 \, ,
\end{equation}
where terms of order $z^5$ are disregarded. For model II, the relative mass
contribution $\mu$ and the prefactor of the $z^3$-term are
\begin{equation} \label{eq:massrenorm2}
\mu_{\rm sc}(s)  = \frac{2 \gamma}{\pi (s-2)}   \, , 
\end{equation}
and
\begin{equation}\label{eq:cubiccoeff2}
\lambda_{\rm sc}(s) = - \mu_{\rm sc} (s-2) =  \frac{2 \gamma}{\pi (4 - s)}     \, .
\end{equation}

\subsection{Mass renormalization}
\label{sec:massrenormalization}

The linear term in $z$ in eqs.~(\ref{eq:gamseriesac}) and
(\ref{eq:gamseriessc}) contributes to the inertial term in the Laplace
transform (\ref{eq:velresplapl}) of the velocity response function. It thus
leads to a mass renormalization which will turn out to be of relevance for the
thermodynamic properties of the free damped particle. We distinguish the
regimes $s<2$ and $s>2$ and start our discussion with the latter. Before, we 
remark though that the motion of a free damped particle in the regime $s>2$
is non-ergodic so that thermodynamic equilibrium is not necessarily being
reached in the long-time limit \cite{Schramm87,Bao05}.

In the parameter regime $s>2$ the linear term in $z$ of $\hat\gamma(z)$ is the
leading one in (\ref{eq:gamseriesac}) and (\ref{eq:gamseriessc}) for $z\ll 1$.
Together with (\ref{eq:gammaHatSpectralDensity}) and (\ref{eq:m_bath}) one
generally finds 
\begin{equation}\label{eq:massrenorm3}
\hat\gamma'(0) = \mu  = \frac{M_\text{bath}}{M}\, ,
\end{equation}
where the prime denotes the derivative with respect to the argument. The
results (\ref{eq:massrenorm1}) and (\ref{eq:massrenorm2}) are special cases of
the general expression (\ref{eq:massrenorm3}). For $s>2$, the dressed mass
$M(1+\mu)$ is larger than the particle's bare mass $M$, since $\mu$
is always positive. The free particle can thus be viewed as dressed by the bath
oscillators. For $s>2$, the appearance of an effective mass is also known for the
ballistic long-time dynamics of the free damped particle \cite{Grabert87}.

\begin{figure}
 \centerline{\includegraphics[width=0.9\columnwidth]{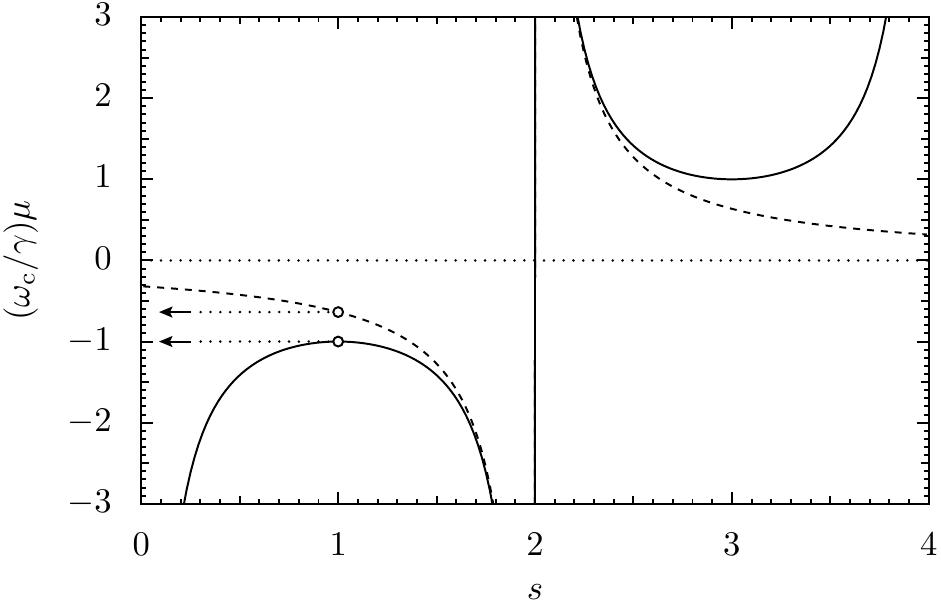}}
 \caption{Mass renormalization $\mu$ as a function of the exponent $s$ of the
 spectral density of the coupling $J(\omega)$.  The solid curve depicts
 $\mu/\gamma$ for an algebraic cutoff with $p=1$ according to
 (\ref{eq:massrenorm1}) while the dashed curve refers to the sharp cutoff, i.e.\
 the expression (\ref{eq:massrenorm2}).  The two small circles indicate how the
 critical value of the damping strength $\gamma^\star$ can be read off for
 $s=1$. For the algebraic cutoff with $p=1$ one finds
 $\gamma^\star_\text{ac}(1,1)=\omega_\text{c}$, while for the sharp cutoff
 $\gamma^\star_\text{sc}(1)=(\pi/2)\omega_\text{c}$.} 
\label{fig:massrenormalization}
\end{figure}

The relative mass renormalization $\mu$ is depicted in
Fig.~\ref{fig:massrenormalization} as a function of the exponent $s$ both for
an algebraic cutoff with $p=1$ (solid line) and a sharp cutoff (dashed line).
For $s>2$, i.e.\ in the right half of the diagram, we notice a significant
difference between the algebraic and the sharp cutoff. In the first case, with
increasing exponent $s$, the mass renormalization goes through a minimum and
diverges at the upper limit of the allowed range (\ref{eq:range_s_p}) at
$2p+2$. For the case $p=1$ presented in the figure, the divergence lies at
$s=4$. In contrast, the mass renormalization for the sharp cutoff decreases
monotonically as $s$ increases. We remark that although for $p=1$ the mass
renormalization for an algebraic cutoff always exceeds the one for the sharp
cutoff, this is generally not true for larger values of $p$.

For $s\leq 2$, the integral (\ref{eq:m_bath}) is infrared-divergent so that the
total mass of the bath oscillators is infinite. Hence the interpretation just
given for the relative mass contribution $\mu$ ceases to hold.  Nevertheless,
the expressions (\ref{eq:massrenorm1}), (\ref{eq:massrenorm2}), and
(\ref{eq:massrenorm3}) can be analytically continued to the regime $0 < s < 2$.
To show this, we write $J(\omega) =J_0(\omega) - [ J_0(\omega) - J(\omega) ]$.
For $s<2$, substitution of the expression (\ref{eq:specdenspowerlaw}) for
$J_0(\omega)$ in (\ref{eq:gammaHatSpectralDensity}) results in the first term
appearing in the series expressions (\ref{eq:gamseriesac}) and
(\ref{eq:gamseriessc}). With the residual contribution of $J(\omega)$, the
integral expression (\ref{eq:gammaHatSpectralDensity}) is regular in linear
order of $z$ and yields $\mu z$ in this order.  Importantly, the mass
contribution $\mu$ emerges negative,
\begin{equation}\label{eq:gamma_cr1}
\begin{aligned}
\mu &= \frac{2}{\pi M} \int_0^\infty\text{d}\omega
\frac{J(\omega)-J_0(\omega)}{\omega^3}\\
&=\frac{M_\text{bath} - M_\text{bath,0}}{M} < 0 \, .
\end{aligned}
\end{equation}
The relation (\ref{eq:gamma_cr1}) unveils that $-\mu M$ represents the total
mass of oscillators which is missing in the actual bath relative to the
reference bath without spectral cutoff \cite{Ingold12,Spreng13}. Hence $\mu$
is negative in the range $0<s<2$, as can also be seen from the left half of
Fig.~\ref{fig:massrenormalization}. Again, we note significant differences
between the algebraic cutoff and the sharp cutoff. While the first one goes
through a minimum in the absolute value of $\mu$, for the latter the
mass renormalization becomes the smallest for $s\to 0$.

As can be expected, the situation where the particle's mass is renormalized
to zero, i.e.\ where $\mu = -1$, is of special physical significance. It
is therefore convenient, to introduce the critical damping strength $\gamma^\star$,
where this point is reached, as
\begin{equation}
1+ \mu = 1  -\frac{\gamma}{\gamma^\star}   \, ,
\end{equation}
and thus
\begin{equation}
\gamma^\star = -\frac{\gamma}{\mu}   \, .
\end{equation}
The two circles and the arrows pointing to the
vertical axis in Fig.~\ref{fig:massrenormalization} indicate, how the inverse
of the critical value $\gamma^\star$ can be read off for $s=1$. The
comparison of the algebraic cutoff with $p=1$ (solid curve) and the sharp cutoff
(dashed curve) indicates that for the latter, a larger damping strength is
required to drive the renormalized mass to zero. Explicit expressions for the
critical damping strength can be obtained for our two models from the expressions
(\ref{eq:massrenorm1}) and (\ref{eq:massrenorm2}) for the relative mass contribution.
For an algebraic cutoff (model I), one obtains
\begin{equation} \label{eq:gamma_cr_model1}
\gamma_\text{ac}^\star   (s,p) = \pi(1-\tfrac{s}{2})\frac{\Gamma(p)}{\Gamma(1+p-\frac{s}{2})
\Gamma(\frac{s}{2})}\,.
\end{equation}
while for the sharp cutoff (model II) the critical damping strength follows as
\begin{equation}\label{eq:gamma_cr_model2}
\gamma_\text{sc}^\star(s) = \pi(1-\tfrac{s}{2})\,.
\end{equation}

\begin{figure}
 \centerline{\includegraphics[width=0.9\columnwidth]{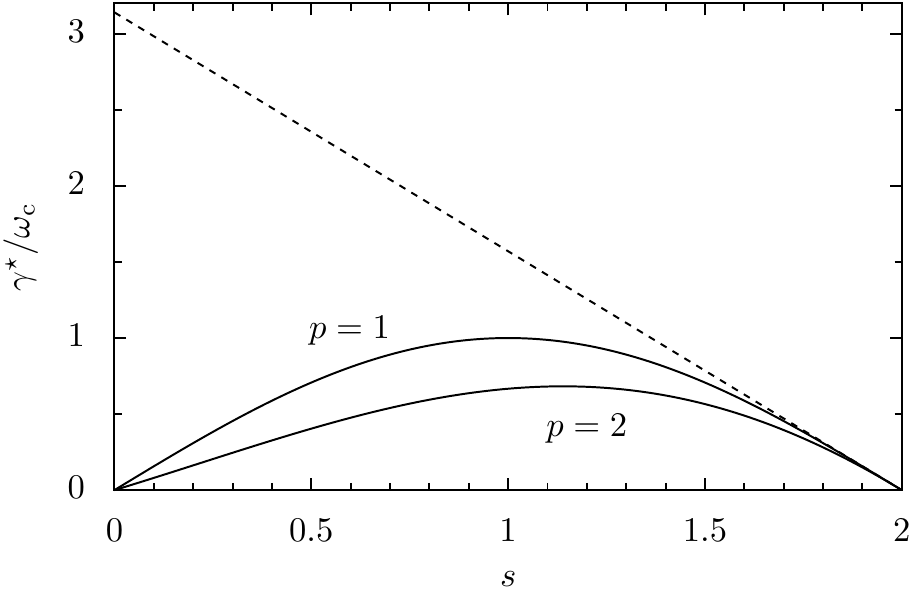}}
 \caption{Critical damping strength $\gamma^\star$ as a function of the
 exponent $s$ in the regime $0<s<2$. The two solid curves show
 $\gamma^\star_\text{ac}$ (model I) for $p=1$ (upper curve) and $p=2$
 (lower curve) while the dashed line depicts $\gamma^\star_\text{sc}$
 (model II).}
\label{fig:critdamping}
\end{figure}

The critical damping strength $\gamma^\star$ is plotted in
Fig.~\ref{fig:critdamping}  both for model I with cutoff parameters $p=1$ and
$p=2$ (solid curves) and for model II (dashed curve). We see that indeed the
critical damping strength for model II is generally larger than for model I.
For Ohmic damping, the critical damping strength is typically of the order of
the cutoff frequency and it is therefore not surprising, that the cutoff
influences the thermodynamic quantities in a significant way. However, when
the exponent $s$ approaches a value of 2 or, in the case of an algebraic
cutoff, approaches zero, the critical damping strength can be much smaller
than the cutoff frequency.

We finally allude again  that the notion of a critical damping strength
$\gamma^\star$ becomes meaningless in the regime $s\ge2$, since there the
analytical continuation, inter alia  of (\ref{eq:gamma_cr_model1}) or
(\ref{eq:gamma_cr_model2}), yields  a negative value for $\gamma^\star$. 

\section{ Change of bath density of states}   
\label{sec:chbds}

\subsection{General considerations}
\label{subsec:gencon}

There are two different views about the specific heat of an open quantum
system: (i) It can be viewed as specific heat of the system modified by the
coupling to the heat bath. (ii) It can be regarded as the change in the
specific heat of the bath when the system degree of freedom is coupled to it
as represented in eq.~(\ref{eq:specheat}). In the latter view which will be
discussed in more detail in Sect.~\ref{sec:redpartfunc_specheat}, the specific
heat of the damped free particle can be expressed in terms of the change of the
density of states of the bath oscillators together with the well-known
expression for the specific heat of a harmonic oscillator.  The change in the
oscillator density of states (CODS) is defined as
\begin{equation}
\xi(\omega) = \sum_{n'=0}^\infty\delta(\omega-\bar\omega_{n'})-\sum_{n=1}^\infty 
\delta(\omega-\omega_n)\,,
\end{equation}
where $\bar\omega_{n'}$ are the eigenvalues of the coupled system-plus-bath
complex (\ref{eq:hamiltonian}), while $\omega_n$ are the eigenfrequencies of
the bath oscillators in the absence of the system-bath coupling, i.e.\ the
frequencies appearing in the second term of the Hamiltonian
(\ref{eq:hamiltonian}).  In the absence of the system-reservoir coupling, the
CODS reduces for the model (\ref{eq:hamiltonian}) to $\xi(\omega)
=\delta(\omega)$.  In the continuum limit of the bath, the density
$\xi(\omega)$ becomes a continuous function of the frequency $\omega$.

The change of the oscillator density of states can be expressed in terms of the
velocity response function (\ref{eq:velresplapl}) as
\begin{equation}
\label{eq:xiexpr1}
\xi(\omega) = \frac{1}{\pi}\text{Im}\frac{\partial
\ln[\hat{\mathcal R}(-\text{i}\omega)]}{\partial\omega}\,,
\end{equation}
where $\text{Im}$ denotes the imaginary part. In terms of the function
\begin{equation}\label{eq:gfunc}
g(\omega) = \frac{\omega-\text{Im}\tilde\gamma(\omega)}
{\text{Re}\tilde\gamma(\omega)} \,,
\end{equation}
where $\tilde\gamma(\omega) = \hat\gamma(-\text{i}\omega)$ is the spectral
damping function in Fourier space, the CODS $\xi(\omega)$ takes the form
\begin{equation}\label{eq:xidef}
\xi(\omega) = \frac{1}{\pi}\frac{g'(\omega)}{1+ g(\omega)^2} 
=  \frac{1}{\pi} \frac{\rm d}{ {\rm d}\omega} \arctan[g(\omega)] \,.
\end{equation}
With the second form, we directly see that the sum rule for the change of the
oscillator density of states reads
\begin{equation}\label{eq:sumrule1}
\Sigma \equiv \int_0^{\infty} \!\!\!\text{d}\omega\,\xi(\omega)   
= \frac{1}{\pi}\big(\arctan[g(\infty)]-\arctan[g(0)]\big)  \,.
\end{equation}
In Sect.~\ref{sec:redpartfunc_specheat} we will see that this sum rule relates
the specific heat in the classical limit with its zero-temperature value.

We will now give expressions for the low-frequency behavior of the
change of the oscillator density of states which pertain to both cutoffs discussed
in the present paper. In the subsequent two subsections, we will take a more
detailed look at each of the two models.

The leading terms of the low frequency series of $g(\omega)$ are found from its
definition (\ref{eq:gfunc}) together with eq.~(\ref{eq:gamseriesac}) for
model~I and eq.~(\ref{eq:gamseriessc}) for model~II as
\begin{equation}\label{eq:gamseries}
g(\omega) = -\cot\left(\frac{\pi s}{2}\right) + \frac{1+\mu}{\gamma} \omega^{2-s}
 - \frac{\Lambda}{\gamma} \omega^{4-s} \, .
\end{equation}
For the models I and II we have, respectively,
\begin{equation}
\label{eq:Lambda}
\begin{aligned}
\Lambda_{\rm ac} &= \lambda_{\rm ac} - p(1+\mu_{\rm ac}) \, , \\
\Lambda_{\rm sc} &= \lambda_{\rm sc} \,.
\end{aligned}
\end{equation}

In the  regime $0<s<2$, the leading terms of the low-frequency series of
$\xi(\omega)$ up to and including $\omega^{3-s}$ are found  from
(\ref{eq:xidef}) with (\ref{eq:gamseries}) as
\begin{equation}\label{eq:xiseries}
\xi(\omega) = \omega^{1-s} \!\left[\frac{2-s}{\pi}\sum_{n=0}^{N_1} b_n
\Big(\frac{1}{\gamma}-\frac{1}{\gamma^\star}\Big)^{n+1} \!\omega^{(2-s)n}
+c\,\omega^2\right]\, .
\end{equation}
Here, the upper limit of summation $N_1$ is given by the largest integer
smaller than $1+1/(2-s)$. Thus, the closer to 2 the parameter $s$ is, the more
terms of the sum must be taken into account. 

The first three coefficients of the first term are
\begin{equation}\label{eq:coefb}
\begin{aligned}
b_0 &= \sin^2(\tfrac{\pi s}{2})\,  ,\\
b_1 &= \sin^2(\tfrac{\pi s}{2})\sin(\pi s)\,,\\
b_2 &= [1+2\cos(\pi s)]\sin^4(\tfrac{\pi s}{2}) \,,
\end{aligned}
\end{equation}
and the coefficient $c$ is
\begin{equation}\label{eq:curvature1}
c =  -\frac{(4-s) \sin^2(\tfrac{\pi s}{2})}{\pi}\, \frac{\Lambda}{\gamma} \, .
\end{equation}
For the models~I and II, $\Lambda$ is given by (\ref{eq:Lambda}).

The leading term proportional to $\omega^{1-s}$ of the series
(\ref{eq:xiseries}) is positive for $\gamma<\gamma^\star$ and negative for
$\gamma > \gamma^\star$.  At critical damping, $\gamma =\gamma^\star$, the
first term in the square brackets in (\ref{eq:xiseries}) vanishes, so that the
leading term for critical damping is given by
\begin{equation}\label{eq:xicritdamp}
\xi(\omega) =  c^\star \omega^{3-s}
\end{equation}
with
\begin{equation}\label{eq:curvcritdamp}
c^\star = - \frac{(4-s) \sin^2(\tfrac{\pi s}{2})}{\pi} \frac{\lambda}{\gamma}\, .
\end{equation}
The ratio $\lambda/\gamma$ in $c^\star$ does not depend on the damping strength
for both models, as we can see from (\ref{eq:coeff3ac}) and
(\ref{eq:cubiccoeff2}).

In the Ohmic case, the $n=2$ term merges with the last term in the series
(\ref{eq:xiseries}) for $\xi(\omega)$ to the curvature contribution at
$\omega=0$. With (\ref{eq:coefb}) and (\ref{eq:curvature1}), we find
\begin{equation} \label{eq:xilfohmic}
\xi(\omega) = \frac{1}{\pi}\left(\frac{1}{\gamma}-\frac{1}{\gamma^\star}\right)-\frac{1}{\pi}
\left[\frac{3\Lambda}{\gamma} + \left(\frac{1}{\gamma}
-\frac{1}{\gamma^\star}\right)^3\right]\omega^2\,. 
\end{equation}

Consider finally the regime $s>2$. Now, the low-frequency expression analogous
to (\ref{eq:xiseries}) with terms of order up to and including $\omega^{s-1}$
reads
\begin{equation}\label{eq:xilfslarge}
\xi(\omega) = \omega^{s-3}\left[-\frac{s-2}{\pi} \sum_{n=0}^{N_2}d_n
\Big(\frac{\gamma}{1+\mu}\Big)^{n+1}\omega^{(s-2)n} + f\omega^2\right] \,.
\end{equation}
The upper limit of summation is given by the largest integer smaller than
$2/(s-2)$ and the coefficients read
\begin{equation}\label{eq:dcoef}
\begin{aligned}
d_0 &= 1\\
d_1 &= 2\cot(\tfrac{\pi s}{2})\\
d_2 &= 3\cot^2(\tfrac{\pi s}{2})-1\,,
\end{aligned}
\end{equation}
and
\begin{equation}\label{eq:fcoef}
f = - \frac{s}{\pi} \frac{\gamma \Lambda}{(1+\mu)^2}  \, .
\end{equation}

For super-Ohmic damping with $s=3$, the $n=2$ term  in (\ref{eq:xilfslarge})
again merges with the last term to the curvature contribution at $\omega=0$.
With the expressions (\ref{eq:dcoef}) and (\ref{eq:fcoef}) we then find 
\begin{equation}
\label{eq:xi_low_super}
\xi(\omega)= -\frac{\gamma}{\pi(1+\mu)} + \frac{\gamma}{\pi} 
\frac{\gamma^2 -3\Lambda(1+\mu)}{(1+\mu)^3 }\omega^2  \,.
\end{equation}

\subsection{Model I}

We now turn to a more specific discussion of the change of the oscillator density of
states for the case of an algebraic cutoff. With the expression
(\ref{eq:gamhatzsmall}) for $\hat \gamma(z)$, the function $g(\omega)$ for
model I is found to read
\begin{equation}
\begin{aligned}
g(\omega) &= -\cot\left(\frac{\pi s}{2}\right) + \frac{1}{\gamma} \omega^{2-s}(1+\omega^2)^p \\
&\quad -\frac{2 p}{\pi (2-s)} {\textstyle B(p+1-\frac{s}{2},\frac{s}{2}) } \\
&\qquad \times\; \omega^{2-s}\, {}_2 F_1(1-p,1-\tfrac{s}{2};2-\tfrac{s}{2}; -\omega^2)\,.
\end{aligned}
\end{equation}
The function $g(\omega)$ is a smooth function of $\omega$ in the range $0\le
\omega<\infty$. It goes to infinity both in the limit $\omega\to\infty$ for all
$s>0$, and in the limit $\omega\to 0$ for $s\ge 2$, whereas $g(\omega= 0) =
\cot(\pi\frac{s}{2})$ for $s<2$. Hence the sum rule (\ref{eq:sumrule1}) reads
\begin{equation}\label{eq:sumrule2} \Sigma(s) = \left( 1-\frac{s}{2}\right)
\,\Theta(2-s) \, .  \end{equation}

The low-frequency expansion of the CODS $\xi(\omega)$ in the regime $0<s<2$
follows from the expressions (\ref{eq:xiseries}), (\ref{eq:coefb}), and
(\ref{eq:curvature1}) together with (\ref{eq:Lambda}), (\ref{eq:massrenorm1})
and (\ref{eq:coeff3ac}).  In the first term in (\ref{eq:xiseries}), the form of
the cutoff enters only through the critical damping strength
$\gamma_\text{ac}^\star$. The leading term $n=0$ in the sum is positive for
$\gamma<\gamma^\star_\text{ac}$ and changes its sign at the critical damping
strength $\gamma^\star_\text{ac}$. The coupling of the system degree of freedom
to the heat bath thus leads to a suppression of the oscillator density if the
damping exceeds $\gamma^\star_\text{ac}$.  The coefficient $c$ in the second
term in (\ref{eq:xiseries}) takes the form
\begin{equation} \label{eq:ccoef_ac}
c_{\rm ac} = \frac{\sin^2(\frac{\pi s}{2})}{\pi}\left[  p(4-s) 
\Big(\frac{1}{\gamma}-\frac{1}{\gamma_{\rm ac}^\star} \Big) +
 \frac{2+2p-s}{\gamma_{\rm ac}^\star}  \right]  \, .
\end{equation}
It is positive for $\gamma <\bar\gamma $ and negative for $\gamma > \bar\gamma$,
where $\bar\gamma = p(4-s)/[(p-1)(2-s)]\gamma_{\rm ac}^\star$.

\begin{figure}
 \centerline{\includegraphics[width=0.9\columnwidth]{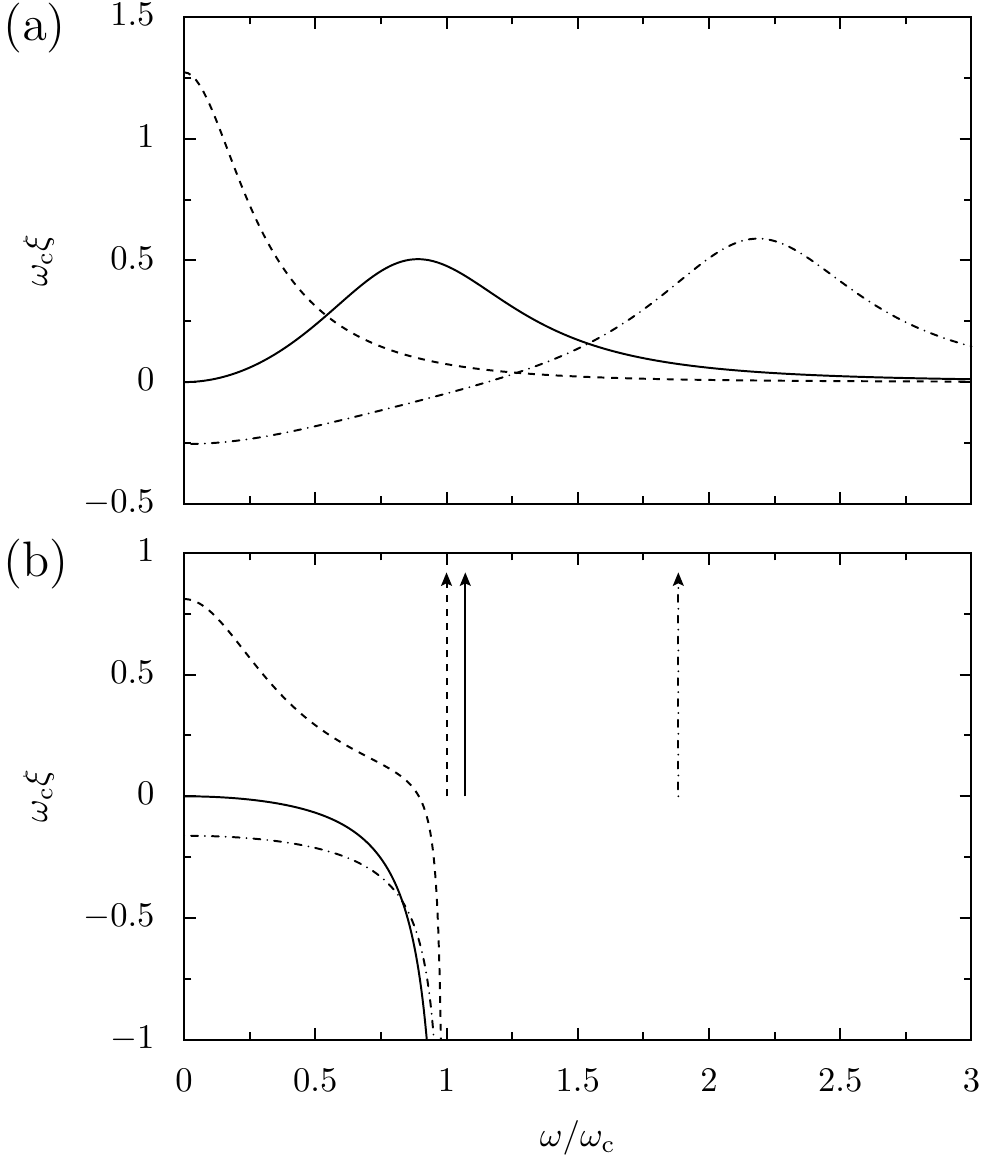}}
 \caption{The change of the oscillator density of states is shown for an Ohmic
 environment (a) with an algebraic cutoff with $p=1$ and (b) with a sharp
 cutoff. The dashed, solid, and dash-dotted curves correspond to
 $\gamma/\gamma^\star=0.2, 1,$ and $5$, respectively. In panel (b), the
 arrows indicate the positions of the delta function in (\ref{eq:denstot}).}
\label{fig:changeofdensity_1}
\end{figure}

For critical damping, $\gamma=\gamma^\star_{\rm ac}$, the coefficient $c^\star$
in the expression (\ref{eq:xicritdamp}) follows from (\ref{eq:ccoef_ac})  as
\begin{equation}\label{eq:coefcaccrit}
c_{\rm ac}^\star(s,p) = \frac{\sin^2(\frac{\pi s}{2})}{\pi}
\frac{2+2p-s}{\gamma_{\rm ac}^\star(s,p)}    \, .
\end{equation}
This coefficient is always positive since the parameters $s$ and $p$ have to
satisfy the condition (\ref{eq:range_s_p}).  In particular, for Ohmic damping
with a Drude cutoff, $s=1$ and $p=1$, we have
\begin{equation}
c_{\rm ac}^\star(1,1) = \frac{3}{\pi}  \, .
\end{equation}
The CODS $\xi(\omega)$ is shown for $s=1$ and $p=1$ in
Fig.~\ref{fig:changeofdensity_1}(a). In this particular case, in which
$\gamma_{\rm ac}^\star = 1$, the low frequency expression (\ref{eq:xilfohmic})
takes the form
\begin{equation}\label{eq:xiohmicac}
\xi(\omega) = \frac{1}{\pi}\left[\frac{1-\gamma}{\gamma}
+\frac{\gamma^3+3\gamma-1}{\gamma^3}\omega^2\right]\,.
\end{equation}
This expression describes the low-frequency behavior of the curves in
Fig.~\ref{fig:changeofdensity_1}(a). While $\xi(0)$ changes its sign at the
critical damping strength, the curvature at $\omega=0$ changes already at the
smaller value $\gamma\approx 0.322\dots$ Correspondingly, the curve for
$\gamma=0.2$ has a negative curvature at $\omega=0$ while the curves for
$\gamma=1$ and $5$ have positive curvature.

For later reference, we note that at low frequencies  the change of the
oscillator density of states for super-Ohmic damping with $s=3$ and an algebraic
cutoff with $p=2$ follows from (\ref{eq:xi_low_super}) as
\begin{equation}
\label{eq:xi_low_super_I}
\xi_\text{ac}(\omega) = -\frac{2\gamma}{\pi(2+\gamma)} +
\frac{48\gamma+12\gamma^2+2\gamma^3}{\pi(2+\gamma)^3}\, \omega^2 \,.
\end{equation}

\subsection{Model II}
\label{subsec:ximodel2}

For the spectral density of the coupling with a sharp cutoff
(\ref{eq:SpecDensSc}), the CODS
$\xi(\omega)$ is a continuous function of $\omega$ in the range $0\le \omega
<1$. For $\omega>1$, the real part of the spectral damping function
$\tilde\gamma(\omega)$ vanishes. Hence the density $\xi(\omega)$ is zero in
this frequency range, except at a particular frequency $\omega=\Omega$, where
the density $\xi(\omega)$ is singular. We thus have
\begin{equation}\label{eq:denstot}
\xi(\omega) = \xi_1(\omega) +\delta(\omega- \Omega) \,,
\end{equation}
where $\xi_1(\omega)$ is the smooth change of the oscillator  density of states in the frequency
range $0\le \omega < 1$,
\begin{equation}
\xi_1(\omega) = \frac{1}{\pi}\frac{g'(\omega)}{1+g(\omega)^2}\Theta(1-\omega)\,.
\end{equation}
Employing the expression (\ref{eq:gamsc}) for $\hat\gamma(z)$, the function
$g(\omega)$ for $0\le\omega<1$ reads
\begin{equation}
\begin{aligned}
g(\omega) &=  -\cot\left(\frac{\pi s}{2}\right) + \frac{1}{\gamma} \omega^{2-s} \\
&\quad -\frac{2}{\pi(2-s)} \omega^{2-s} \,
{}_2 F_1(1,1-\tfrac{s}{2};2-\tfrac{s}{2} ; \omega^2) \, .
\end{aligned}
\end{equation}
With this particular form, the sum rule for the partial  density
$\xi_1(\omega)$ is found as 
\begin{equation}
\Sigma_1(s) = - \frac{s}{2}\,\Theta(2-s)- \Theta(s-2) \, .
\end{equation}

We see from the general expression (\ref{eq:xidef}) with (\ref{eq:gfunc}) that
the frequency $\Omega>1$, at which the function $\xi(\omega)$ is singular, is a
zero of the function
\begin{equation}
\begin{aligned}
N(\omega,s) &\equiv \omega- {\rm Im}\,\tilde\gamma(\omega) \\
&= \omega -\frac{2\gamma}{\pi s}\, 
\frac{ {}_2F_1(1,\frac{s}{2},1+\frac{s}{2},\frac{1}{\omega^2})}{\omega} \,  .
\end{aligned}
\end{equation}
In the particular cases $s=1$ and $s=3$ there holds
\begin{equation}
\begin{aligned}
N(\omega,1) &= \omega-\frac{2\gamma}{\pi}\text{arcoth}(\omega)  \, ,\\
N(\omega,3) &= \omega\left(1 +\frac{2\gamma}{\pi} \right) - \frac{2\gamma}{\pi}
\omega^2\text{arcoth}(\omega) \, .
\end{aligned}
\end{equation}

The singular term in (\ref{eq:denstot}) contributes unity to the integrated
change of the oscillator density of states. Hence the sum rule for the total change of the
density of bath oscillators (\ref{eq:denstot}) is
\begin{equation}\label{eq:cbosc}
\Sigma(s) = \Sigma_1(s)+1 = \left(1-\frac{s}{2}\right)\Theta(2-s) \, ,
\end{equation}  
which is in full agreement with the corresponding expression
(\ref{eq:sumrule2}) of model I.

The explicit expression for the low-frequency expansion of the change of the
oscillator density (\ref{eq:xiseries}) for model~II follows from
(\ref{eq:coefb}) and (\ref{eq:curvature1}) together with (\ref{eq:Lambda}),
(\ref{eq:massrenorm2}) and (\ref{eq:cubiccoeff2}).  As for model~I, the leading
term $n=0$ in the expression (\ref{eq:xiseries}) is positive for
$\gamma<\gamma^\star$ and negative for $\gamma >\gamma^\star$.  The coefficient
$c$ does not depend on $\gamma$ at all and is found to read
\begin{equation}\label{eq:b3coefsc}
c_{\rm sc}(s)=c_{\rm sc}^\star (s)= -\frac{2}{\pi^2}\sin^2(\tfrac{\pi s}{2}) \, .
\end{equation}

While the first term in (\ref{eq:xiseries}) for both models depends only on the
form of the cutoff through the critical damping strength $\gamma^\star$, the
coefficient $c^\star$ in the expression (\ref{eq:xicritdamp}) behaves
qualitatively different for the two models. The coefficient $c^\star_{\rm ac}$
of model~I given in (\ref{eq:coefcaccrit}) is generally positive, whereas the
coefficient $c_{\rm sc}^\star$  of model~II is generally negative.

The CODS $\xi(\omega)$  for model~II is shown in
Fig.~\ref{fig:changeofdensity_1}(b) for Ohmic damping, $s=1$.  In this case,
the critical damping strength is $\gamma_{\rm sc}^\star =\pi/2$, and the
expression  (\ref{eq:xilfohmic})  takes the form
\begin{equation}\label{eq:xiohmicsc}
\xi(\omega) = \frac{1}{\pi} \Big(\frac{1}{\gamma} -\frac{1}{\gamma_{\rm sc}^\star}\Big)
- \frac{1}{\pi}\left[\frac{1}{\gamma_{\rm sc}^\ast} +   
\Big(\frac{1}{\gamma} -\frac{1}{\gamma_{\rm sc}^\star}\Big)^3 \right]\omega^2 \, .
\end{equation}
In contrast to the expression (\ref{eq:xiohmicac}) for the algebraic cutoff,
the curvature at small frequencies is negative for arbitrary damping strength
$\gamma$.  The different characteristics of $\xi(\omega)$ at low frequencies
displayed in Figs.~\ref{fig:changeofdensity_1}(a) and (b) are reproduced by the
expressions (\ref{eq:xiohmicac}) and (\ref{eq:xiohmicsc}). 

\begin{figure}
 \centerline{\includegraphics[width=0.9\columnwidth]{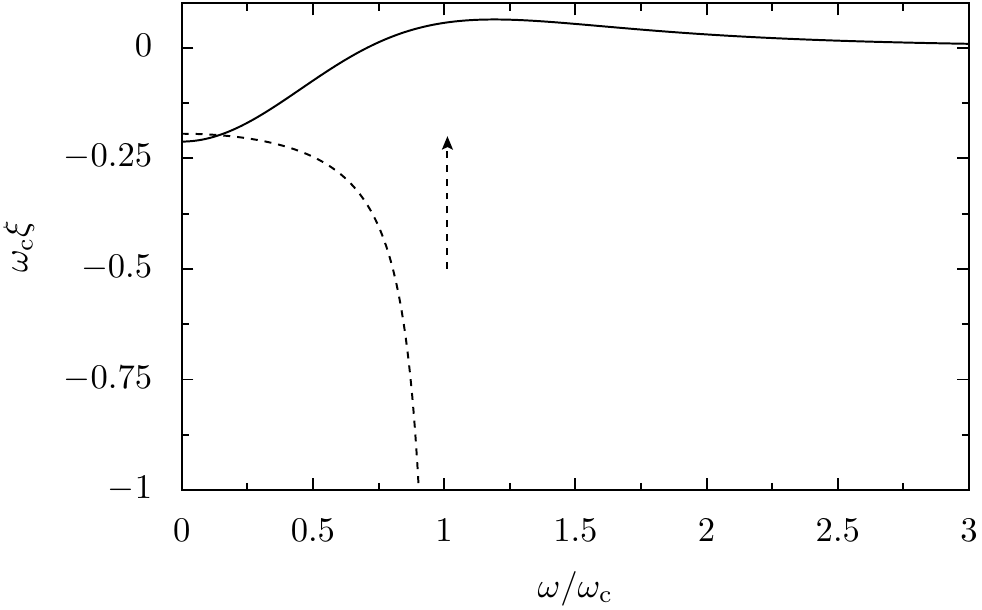}}
 \caption{The change of the oscillator density of states is shown for a
 super-Ohmic environment with $s=3$ and $\gamma=\omega_\text{c}$ with
 an algebraic cutoff with $p=2$ (solid curve) and with a sharp cutoff
 (dashed curve). The arrow indicates the position of the delta function
 in (\ref{eq:denstot}).}
\label{fig:changeofdensity_3}
\end{figure}

The difference in  curvature of the function $\xi(\omega)$ at zero frequency in
models~I and II also prevails for super-Ohmic damping with $s=3$, as is clearly
visible in Fig.~\ref{fig:changeofdensity_3}. The qualitative difference between
model~I (solid curve) and model~II (dashed curve) in
Fig.~\ref{fig:changeofdensity_3} can be understood from the leading terms of
the low-frequency series of $\xi(\omega) $ for $s=3$ given in
(\ref{eq:xi_low_super}). The low-frequency expansion of the change of the oscillator
density of states for model~I is given in (\ref{eq:xi_low_super_I}). For
model~II, we obtain from (\ref{eq:xi_low_super}) together with
(\ref{eq:massrenorm2}), (\ref{eq:cubiccoeff2}) and (\ref{eq:Lambda})
\begin{equation}\label{eq:xisuperohmic}
\xi_\text{sc}(\omega) = - \frac{\gamma}{\pi +2\gamma}
-\gamma^2 \frac{6\pi +(12-\pi^2)\gamma}{(\pi+2\gamma)^3}\omega^2\,.
\end{equation}
Importantly, the different sign of the curvature in models~I and II
displayed in (\ref{eq:xi_low_super_I}) and (\ref{eq:xisuperohmic}), respectively,
holds for general damping strength $\gamma$.

\section{Reduced partition function and specific heat}
\label{sec:redpartfunc_specheat}

Within the present paper, we determine thermodynamic quantities of open systems
on the basis of the reduced partition function
(\ref{eq:reduced_partition_function}). For a free particle, the partition
function is only well defined if the particle is confined to a finite region.

In the absence of an environmental coupling, we can evaluate the partition
function of a particle confined to a one-dimensional infinite square well of
width $L$ and inner potential $V_0=0$. The corresponding eigenenergies are
given by $E_n = E_\text{g}n^2$ with $E_\text{g}=\pi^2/2ML^2$ so that the
partition function reads
\begin{equation}
 {\mathcal Z}_0 = \sum_{n=1}^\infty\exp\left(-\frac{E_\text{g}}{T}n^2\right)
= \frac{1}{2}\left[\vartheta_3\big(0, {\rm e}^{-E_{\rm g}/T}\big) - 1\right]\,.
\end{equation}
Here, $\vartheta_3(0,x)$ is a Jacobi theta function \cite{NISTHandbook10}. In
the temperature regime
\begin{equation}
\label{eq:temp0}
 T > E_0 = c E_{\rm g} \, ,
\end{equation}
where $c$ is a positive number sufficiently large so that for $T$ above
$E_0$ the discreteness of the energy eigenstates may be disregarded, the sum
in the partition function can be turned into an integral. We thus arrive
at the classical partition function of the undamped free particle, 
\begin{equation}
\label{eq:zetnull}
\mathcal{Z}_{0,{\rm cl}} = \sqrt{\frac{T}{E_\text{g}}}\int_0^\infty\text{d}x
\exp(-x^2) =\sqrt{\frac{\pi T}{4E_\text{g}}}\, .
\end{equation}
This result depends only on the combination $TL^2$ and therefore is valid
even for very low temperatures provided the particle is constrained to a
sufficiently large region. How the confinement to a finite region influences
the thermodynamic properties at very low temperatures, can for example be
seen by studying a damped particle in a weakly confining harmonic potential
\cite{Adamietz14}.

For a damped particle, the partition function is augmented by quantum
fluctuations due to the bath coupling. The accessory part may be written as an
infinite Matsubara product \cite{Weiss12} which, under the condition
(\ref{eq:temp0}), does not depend on the width $L$ of the square well. The
resulting reduced partition function reads
\begin{equation}
\label{eq:partfunc}
\mathcal{Z} = \sqrt{\frac{\pi\,T}{4 E_{\rm g}}}\,\, \prod_{n=1}^\infty \frac{\nu_n}{\nu_n +\hat\gamma(\nu_n)} \, , \qquad T>E_0 \, ,
\end{equation}
in which $\nu_n =2\pi T n $ is a bosonic Matsubara frequency.
The subsequent thermodynamic analysis is based on the expression
(\ref{eq:partfunc}).

The specific heat follows from the reduced partition function by
\begin{equation}
\label{specheat1}
C = \frac{\partial}{\partial T}\left(T^2\frac{\partial\ln(\mathcal{Z})}{\partial T}
\right)\,.
\end{equation}
Based on the Matsubara representation (\ref{eq:partfunc}), one finds
\begin{equation}
\label{eq:specheatmatsu}
C = \frac{1}{2}+\sum_{n=1}^{\infty} y(\nu_n)
\end{equation}
with the function \cite{Spreng13}
\begin{equation}
\label{eq:matsubarafunction}
y(\nu) = \left(\frac{\hat\gamma(\nu)-\nu\hat\gamma'(\nu)}{\nu+\hat\gamma(\nu)}\right)^2
   -\frac{\nu^2\hat\gamma''(\nu)}{\nu+\hat\gamma(\nu)} \, ,
\end{equation}
where the prime denotes again the derivative.

The Matsubara series (\ref{eq:specheatmatsu}) is advantageous for
moderate-to-high temperatures. From it we can immediately infer that $C$ tends
to $1/2$ as $T\to\infty$. Regrettably, it is badly converging at low
temperatures. In the latter regime, it is pertinent to perform a Poisson
resummation of the series (\ref{eq:specheatmatsu}) by virtue of the
periodically continued $\delta$-function 
\begin{equation}
{:}\delta(\tau){:} = T \sum_{n=-\infty}^\infty \exp({\rm i}\nu_n \tau)
\end{equation}
into the form
\begin{equation}
\label{eq:specheatresummed}
\begin{aligned}
C &= \frac{1}{2}-\frac{y(0)}{2}+\frac{1}{2\pi T}\int_0^\infty\text{d}\nu \,y(\nu)\\
&\qquad+\frac{1}{\pi T}\sum_{n=1}^\infty\int_0^\infty\text{d}\nu\cos\left(n\frac{\nu}{T}\right) y(\nu) \, .
\end{aligned}
\end{equation}
Since the antiderivative
\begin{equation}
Y(\nu) = \frac{\nu [\hat\gamma(\nu) - \nu \hat\gamma'(\nu)]}{\nu+\hat\gamma(\nu)}\,  .
\end{equation}
of the function $y(\nu)$ vanishes at the boundaries $\nu=0$ and $\nu =\infty$,
the first integral in (\ref{eq:specheatresummed}) is strictly zero.  In
addition, the trigonometric function in the second integral of
(\ref{eq:specheatresummed}) is increasingly oscillating as $T\to 0$. Hence the
second integral approaches zero in this limit. Thus the specific heat
(\ref{eq:specheatresummed}) can be expressed in the form
\begin{equation} \label{eq:specheat1}
C = C_0 +  \frac{1}{\pi T}\sum_{n=1}^\infty\int_0^\infty\text{d}\nu\cos\left(n\frac{\nu}{T}\right) y(\nu) \, , 
\end{equation}
where
\begin{equation}\label{eq:sphzero0}
C_0 = \frac{1}{2}-\frac{y(0)}{2}  
\end{equation}
is the specific heat at zero temperature.  We find from
(\ref{eq:matsubarafunction}) both with the form (\ref{eq:gamseriesac}) and  the
form (\ref{eq:gamseriessc}) for the spectral damping function the limiting
expression 
\begin{equation}\label{eq:fzero}
 y(0)=(2-s)\Theta(2-s)\, . 
\end{equation}
The  second term in  (\ref{eq:specheat1}) describes the temperature dependence
of the specific heat and can serve to obtain low-temperature expansions.

A more physical interpretation of the specific heat for a damped system can be
given in terms of the change $\xi(\omega)$ of the oscillator density of states caused by
the coupling of the system to the heat bath \cite{Ingold12},  which has been
discussed in Sect.~\ref{sec:chbds}. Under the condition
(\ref{eq:range_s_p_restricted}) the Matsubara sum (\ref{eq:specheatmatsu}) can
be rewritten as the frequency integral
\begin{equation}
\label{eq:specheatspectral}
\begin{aligned}
C &= \frac{1}{2} + \int_0^\infty {\rm d}\omega\, \xi(\omega)
           \left[C_{\rm ho}(\omega)-1\right]\, ,  \\
&=  C_0 + \int_0^\infty {\rm d}\omega\, \xi(\omega) \,
           C_{\rm ho}(\omega)\, .
\end{aligned}
\end{equation}
Here,
\begin{equation}
\label{harmosc}
C_{\rm ho}(\omega) = \left(\frac{\omega}{2T\sinh(\omega/2T)}\right)^2 
\end{equation}
is the  specific heat of a single harmonic oscillator with frequency
$\omega$. Since in the high-temperature limit, $C_\text{ho}$ tends to one,
the first term on the right-hand side of the first line of (\ref{eq:specheatspectral})
represents the classical value of the specific heat. On the other hand, in
the low-temperature limit, $C_\text{ho}$ tends to zero, thus confirming that
$C_0$ is the specific heat at zero temperature. By comparing the first
and the second line, we find that the specific heat in the zero-temperature limit
\begin{equation}\label{eq:sphzero1}
C_0 = \frac{1}{2} - \Sigma 
\end{equation}
is related to its classical value by the integrated CODS $\Sigma$
introduced in (\ref{eq:sumrule1}).

With the explicit form (\ref{eq:sumrule2}) for the sum rule $\Sigma$, the
expression (\ref{eq:sphzero1}) in fact coincides with the former result
(\ref{eq:sphzero0}) with (\ref{eq:fzero}). In any event, the resulting
expression for the specific heat at zero temperature is
\begin{equation}
\label{eq:specheatzerotemp}
C_0 = \begin{cases} \dfrac{s-1}{2} &\qquad \text{for $s\le2$}\, ,\\[4mm]
\quad \dfrac{1}{2} &\qquad \text{for $s>2$\, .}\end{cases}
\end{equation}
In the regime $s\le2$, $C_0$ increases linearly with the exponent $s$ from
a value of $-1/2$ in the extreme sub-Ohmic regime to the classical value
of $1/2$ which is reached for $s=2$. 

The expressions for the specific heat given in (\ref{eq:specheatspectral})
differ from the expression which can be derived from the formula for the free
energy given by Ford et al.\ \cite{Ford85,Ford88} for the damped harmonic
oscillator. The reason lies in the nonvanishing specific heat of the free
damped particle at zero temperature.

\begin{figure}
 \centerline{\includegraphics[width=0.9\columnwidth]{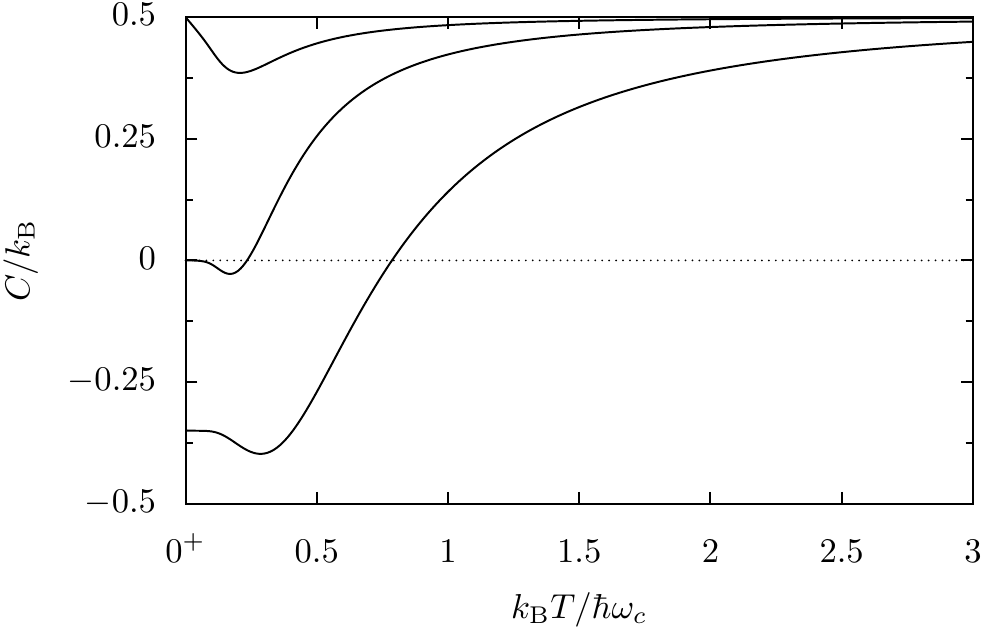}}
\caption{The specific heat (\ref{eq:specheatmatsu}) of a free damped particle
 is displayed as a function of temperature for spectral densities of the
 coupling (\ref{eq:SpecDensSc}) with a sharp cutoff. The label $0^+$ on the
 temperature axis indicates that the curves are only valid for
 $k_\text{B}T>E_0$. The exponent $s$ increases from the lower to the upper curve
 as $s=0.3, 1,$ and $3$. For the two smaller values of $s$, the critical damping
 strength $\gamma=\gamma^\star_\text{sc}$ is chosen while for $s=3$ the damping
 strength is set to $\gamma=1$.}
 \label{fig:specificheat}
\end{figure}

Figure~\ref{fig:specificheat} gives an overview of typical variations of the
specific heat of a damped free particle as a function of temperature. In order
to emphasize that the represented data are only valid as long as the condition
(\ref{eq:temp0}) is satisfied, i.e.\ provided that $TL^2$ is sufficiently large,
we denote the leftmost value on the temperature axis by $0^+$. The three curves
represent a sub-Ohmic case ($s=0.3$), the Ohmic case ($s=1$) and a super-Ohmic
case ($s=3$) from the lower to the upper curve. In all cases, a sharp cutoff
(model~II) has been employed.

As our previous analysis has shown, all three curves lead from a
zero-temperature value of the specific heat given by
(\ref{eq:specheatzerotemp}) to the classical value $k_\text{B}/2$. The Ohmic
case is a particular case because it leads to a vanishing specific heat in the
zero-temperature limit independently of any confinement condition for the
particle. In the sub-Ohmic case, the specific heat at low temperatures tends
towards a negative value. 

\begin{figure}
 \centerline{\includegraphics[width=0.9\columnwidth]{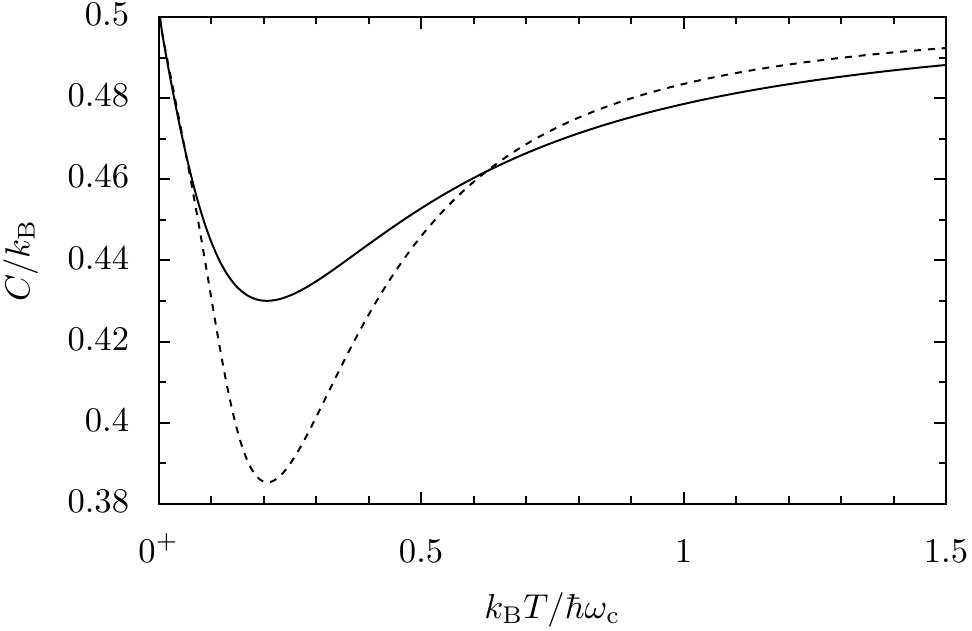}}
 \caption{The specific heat is shown for super-Ohmic damping $s=3$ with an
 algebraic cutoff with $p=2$ (solid curve) and a sharp cutoff (dashed curve).
 The damping strength for both curves is $\gamma=1$. The dashed curve is a
 magnified form of the topmost curve in Fig.~\ref{fig:specificheat}. We see
 that the dip in the specific heat is notably larger for model~II compared
 with model~I.}
 \label{fig:specheatsequ3}
\end{figure}

In contrast, for super-Ohmic damping with $s>2$, which includes the case $s=3$
represented in Fig.~\ref{fig:specificheat}, at low temperatures the classical
value of the specific heat is approached again. This phenomenon of reentrant
classicality is due to the small density of bath oscillators at low frequencies
\cite{Spreng13}. In Fig.~\ref{fig:specheatsequ3} the reentrant classicality for
the super-Ohmic case $s=3$ is shown both for model~I and model~II.  The dip in
the specific heat due to quantum effects is considerably larger in the dashed
curve belonging to model~II compared with  the solid curve pertaining to
model~I. The discriminative characteristics in Fig.~\ref{fig:specheatsequ3} is
due to the drastically different behaviors of the CODS $\xi(\omega)$
of these models shown in Fig.~\ref{fig:changeofdensity_3}. It should be
remarked, that the dip for both models gets deeper, as the parameter $\gamma$
is increased, while the characteristics of reentrant classicality is preserved.

As far as the two lower curves in Fig.~\ref{fig:specificheat} belonging to the
range $0<s<2$ are concerned, their dips get  progressively deeper, as the
damping strength $\gamma$  is increased beyond $\gamma^\star$ \cite{Spreng13}.
This feature is due to the negative sign of the leading term in the expansion
(\ref{eq:xiseries}) in the regime  $\gamma>\gamma^\star$.

The overall structure discussed so far only depends on the exponent $s$ of the
spectral density of the coupling and is thus independent of the cutoff. The
cutoff becomes relevant though at finite temperatures. Accessible to a
analytical analysis are the leading quantum corrections at high temperatures
which we will discuss in Sect.~\ref{sec:hightemp} and the low-temperature
expansion which will be the subject of Sect.~\ref{sec:lowtemp}. The low-temperature
behavior is particularly interesting because for sufficiently strong coupling of
the free particle to its environment, the specific heat at finite temperatures can
fall even below its zero-temperature value.

\section{Quantum corrections at high temperatures}
\label{sec:hightemp}

The specific heat at temperatures $T\gg \gamma,\,\omega_\text{c}$ is determined
by the behavior of the Laplace transform $\hat\gamma(z)$ of the damping kernel
in the regime $|z|\gg 1$, which is given in eq.~(\ref{eq:gamht}). In the
high-temperature regime, the Matsubara sum (\ref{eq:specheatmatsu}) is
dominated by the second term in (\ref{eq:matsubarafunction}). As the damping
kernel decays like $1/z$ for all exponents $s$ in the range
(\ref{eq:range_s_p_restricted}), the leading quantum correction at high
temperatures then goes like the square of the inverse temperature.
\begin{equation}
\label{eq:sphhtl}
C = \frac{1}{2} - \frac{a}{12}\frac{1}{T^2}  \, .
\end{equation}
The coefficient $a$ is given in (\ref{eq:acoef1}) for model I and in
(\ref{eq:acoef2}) for model II.
 
\begin{figure}
 \centerline{\includegraphics[width=0.9\columnwidth]{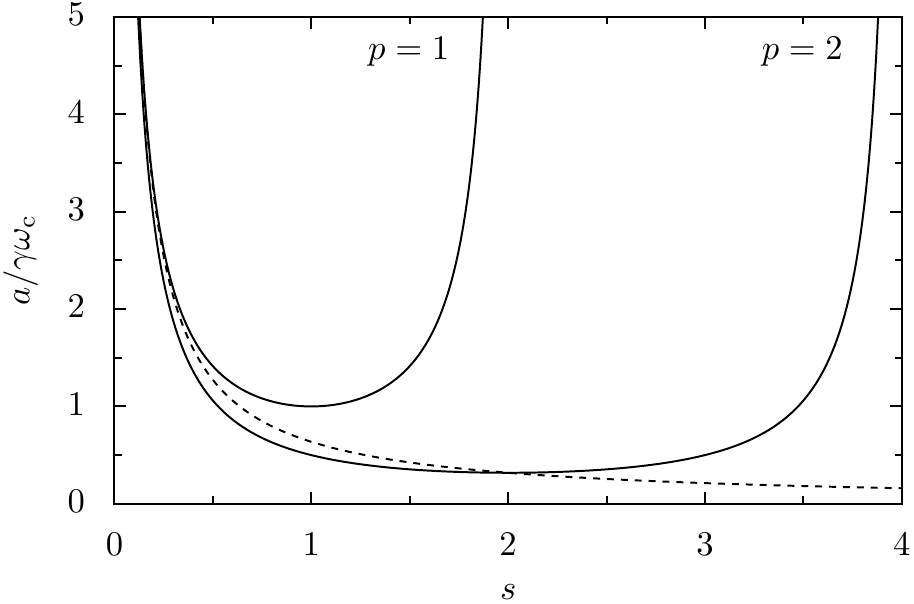}}
 \caption{The amplitude functions $a_{\rm ac}(s,p)$ for algebraic cutoffs
 characterized by exponents $p=1$ and $p=2$ (solid curves) and the amplitude
 function $a_{\rm sc}(s)$ for sharp cutoff (dashed curve) appearing in the
 high-temperature formula (\ref{eq:sphhtl}) are shown as a function of the
 exponent $s$.}
 \label{fig:htlcf}
\end{figure}

The universal $1/T^2$ tail is proportional to the damping strength $\gamma$
and, after reinserting the constants previously set to one, the cutoff
frequency $\omega_c$. The amplitude functions $a_{\rm ac}(s,p)/\gamma$ for
model I and $a_{\rm sc}(s)/\gamma$ for model II are plotted in
Fig.~\ref{fig:htlcf}. The  U-shaped form for model I (displayed for $p=1$ and
$p=2$) possesses a minimum at $s=p$ and divergencies at the edges $s=0$ and
$s=2p$. For a given algebraic cutoff function characterized by $p$, quantum
effects can depend significantly on the value of the exponent $s$. They are
weakest for $s=p$.

When, on the other hand, $s$ is kept fixed, the function $a_{\rm ac}(s,p)$
decreases with increasing parameter $p$. Thus, sharpening the cutoff function
in the spectral density of the coupling
(\ref{eq:spectralDensityOfBathOscillators}) in model I reduces the quantum
effects while the temperature is kept fixed at a large value.

For the sharpest possible cutoff function, i.e.\ our model~II, the amplitude
$a_\text{sc}$ decreases monotonically as $1/s$ with increasing exponent $s$.
Although, the sharp cutoff does in general not lead to the smallest
high-temperature quantum corrections, it nevertheless fits the general picture
that sharper cutoffs in the spectral density of the coupling tend to lead to
weaker quantum corrections.

\section{Low-temperature behavior}
\label{sec:lowtemp}

We finally study the low temperature regime where $T$ provides the smallest
frequency scale, i.e.\ $T\ll\gamma, \omega_\text{c}$. As mentioned before, for
a particle confined to a finite spatial region, our results are constrained by
the condition (\ref{eq:temp0}) and are thus not valid down to zero temperature.
However, by choosing the confining region sufficiently large, the domain of
validity can be extended down to any arbitrarily low non-zero temperature. The
low-temperature expansions for the specific heat given in this section have to
be understood in this sense.

The leading low-temperature correction to the specific heat is obtained from
the leading low-frequency term in the change of the oscillator density of states
$\xi(\omega)$ discussed in Subsection \ref{subsec:gencon}.  In the regime
$0<s<2$, the leading contribution to the CODS $\xi(\omega)$ is found to be
proportional to $\omega^{1-s}$. From the expression
for the specific heat in the second line of (\ref{eq:specheatspectral}), we
obtain the leading low-temperature behavior of the specific heat as
\begin{equation}
\begin{aligned}
\label{eq:sphlt}
C &= \frac{s-1}{2}\\
&\quad +  \frac{2-s}{\pi}\Big(\frac{1}{\gamma}-\frac{1}{\gamma^\star}\Big)
\sin^2(\tfrac{\pi s}{2})\Gamma(4-s)\zeta(3-s)T^{2-s}\,.
\end{aligned}
\end{equation}
Here, $\zeta$ denotes the Riemann zeta function \cite{NISTHandbook10}. The
expansion (\ref{eq:sphlt}) depends mainly on the low-frequency dependence
of the spectral density of the coupling. Its dependence on the form of the
cutoff enters only via the critical damping strength $\gamma^\star$.

The leading low-temperature correction to the specific heat increases with
decreasing damping strength $\gamma$.  Consequently, a reduction of the
environmental coupling leads to a more rapid approach to the classical regime
as temperature is increased. This behavior is consistent with the fact that in
the absence of any spatial confinement, the environmental coupling provides the
mechanism to render a free particle quantum mechanical
\cite{Haenggi06,Haenggi08,Spreng13}.

\begin{figure}
 \centerline{\includegraphics[width=0.9\columnwidth]{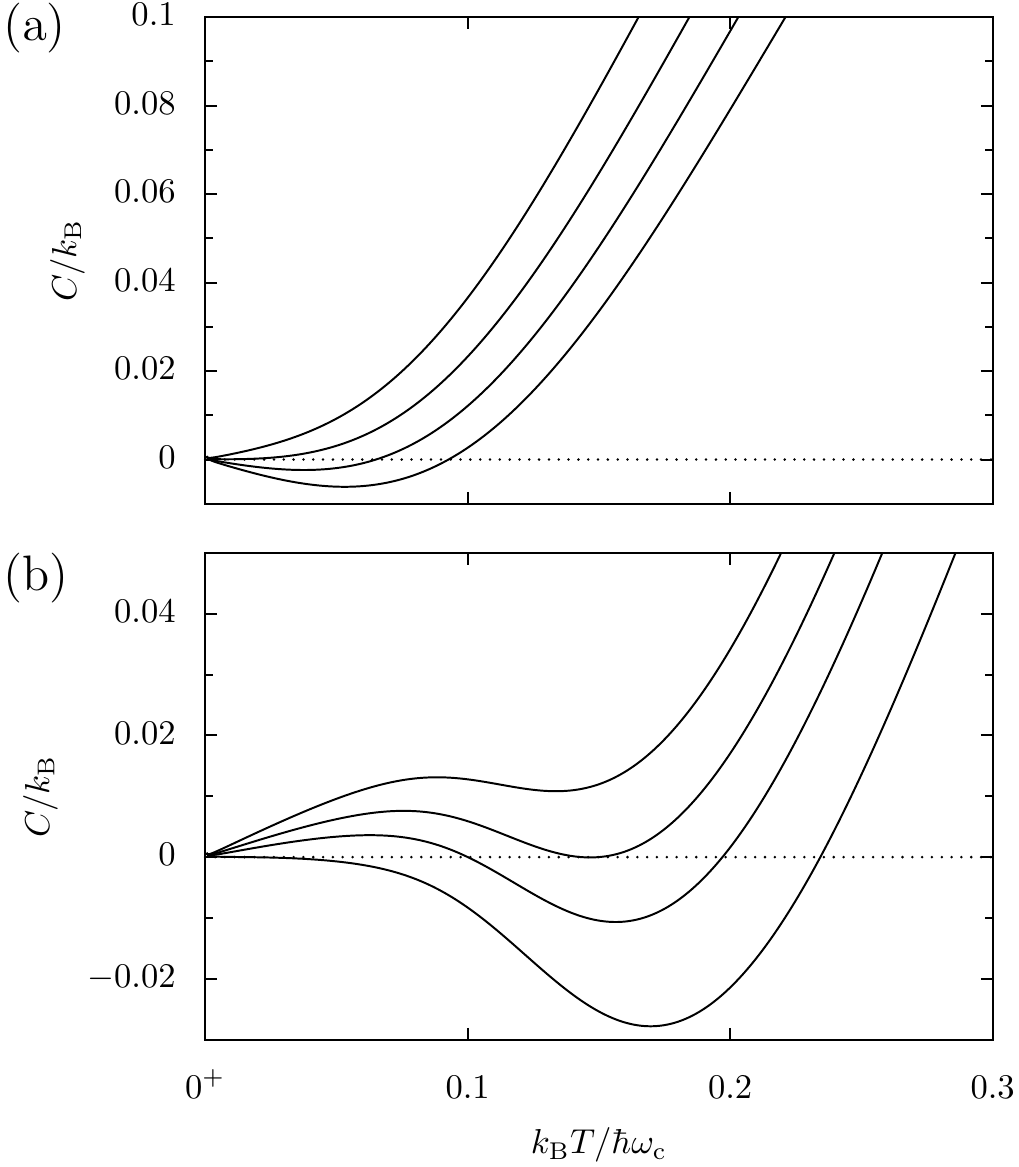}}
 \caption{The specific heat at low temperatures where negative values can occur
 is shown for Ohmic damping with (a) an algebraic cutoff with $p=1$ and (b) a
 sharp cutoff. In the upper panel, the damping strength $\gamma/\omega_\text{c}
 = 0.9, 1, 1.1$, and $1.2$ increases from the upper to the lower curve. The
 second value represents the critical damping strength $\gamma=
 \gamma^\star_\text{ac}$.  In the lower panel, the damping strength takes the
 values $\gamma/\omega_\text{c} = 1.2, 1.3, 1.4$, and $\pi/2$ from the upper to
 the lower curve. The last choice corresponds to $\gamma =
 \gamma^\star_\text{sc}$. We see that already for values $\gamma <
 \gamma^\star_\text{sc}$ the specific heat can become negative at finite
 temperatures.} 
 \label{fig:c_lowt_1}
\end{figure}

With increasing damping strength, eventually the critical damping strength
$\gamma^\star$ will be reached where the dressed mass $M(1+\mu)$ vanishes.
According to (\ref{eq:xiseries}), the leading term of $\xi(\omega)$ then
changes sign and the leading low-temperature correction results in a decrease
of the specific heat with increasing temperature. In
Fig.~\ref{fig:c_lowt_1}(a), we show the low-temperature behavior of the
specific heat for Ohmic dissipation with an algebraic cutoff with $p=1$.  The
damping strength in Fig.~\ref{fig:c_lowt_1}(a) increases from the upper to the
lower curve, clearly demonstrating the change in the sign of the leading term
in the CODS $\xi(\omega)$ at the critical damping strength
represented by the second curve from the top. 

Interestingly, the low-temperature behavior for a sharp cutoff shown in
Fig.~\ref{fig:c_lowt_1}(b) is qualitatively different. While we see again how
the leading term in (\ref{eq:xiseries})  changes its sign when the critical
damping strength is reached at the lowest curve, a negative specific heat can
be obtained even for damping strengths below the critical damping strength.
However, the specific heat falls below its zero-temperature value only above a
certain finite temperature.

This difference in behavior of the specific heat for algebraic and sharp cutoff
can be traced back to the coefficient $c$ in the expansion (\ref{eq:xiseries}).
This coefficient dominates the low-temperature behavior of the specific heat at
the critical damping strength $\gamma=\gamma^\star$ where the first term in
(\ref{eq:xiseries}) vanishes. The specific heat then reads 
\begin{equation} 
C= \frac{s-1}{2} + c^\star \, \Gamma(6-s)\zeta(5-s)T^{4-s}\, , 
\end{equation}
where the coefficient $c^\star$ is given in (\ref{eq:coefcaccrit}) for model~I and
in (\ref{eq:b3coefsc}) for model~II. Recalling the discussion at the end of
Subsection \ref{subsec:ximodel2}, the leading thermal contribution for critical
damping $\gamma=\gamma^\star$ is generally positive for model~I and generally
negative for model~II. For a sharp cutoff, the absolute value of the
coefficient $c$ can be large enough, even for undercritical damping $\gamma
<\gamma^\star$,  to force the specific heat below zero as can be seen in
Fig.~\ref{fig:c_lowt_1}(b).

In the regime $s>2$, we had seen from (\ref{eq:specheatzerotemp}) that at zero
temperature, the specific heat takes it classical value. Upon using the
expression (\ref{eq:xilfslarge}) for the change of the oscillator  density of states
$\xi(\omega)$, we find that the leading contribution to the specific heat at
low temperatures is
\begin{equation}
\label{eq:clt_rc}
C = \frac{1}{2} - (s-2) \frac{\gamma}{\pi(1+\mu)}\Gamma(s)\zeta(s-1) T^{s-2}\,.
\end{equation}
Interestingly, details of the cutoff in the spectral density of the coupling
$J(\omega)$ only enter via the mass renormalization $\mu$. The leading thermal
contribution in (\ref{eq:clt_rc}) is always negative, thereby ensuring that the
specific heat never exceeds its classical value.

\section{Conclusions}
\label{sec:conclusions}

The thermodynamic properties of a damped quantum system depend on the spectral
density of the coupling, in particular its low-frequency behavior and the
high-frequency cutoff. Choosing a rather general spectral density of the form
(\ref{eq:specdens}), we have seen for the specific heat of the free damped
particle, that the exponent $s$ characterizing the low-frequency properties
of the environmental coupling is omnipresent. Nevertheless, also the existence
of the high-frequency cutoff and its detailed form are of relevance.

First of all, the negative mass renormalization in the regime $0<s<2$ is a 
consequence of the mere existence of a high-frequency cutoff. In this sense,
the possibility for the specific heat to fall below its zero-temperature value
constitutes an effect of the cutoff. Typically, this requires that damping
strength and cutoff frequency are of the same order which makes the appearance
of cutoff effects likely. However, as we have seen in Fig.~\ref{fig:critdamping},
there exist parameter ranges where even relatively weak damping can lead
to a decrease of the specific heat at low temperatures.

A qualitative difference  between the two types of cutoffs considered here,
i.e.\ the algebraic cutoff (\ref{eq:powerlawSpectralDensity}) and the sharp
cutoff (\ref{eq:SpecDensSc}), can be found in the change of the oscillator
density of states when the system degree of freedom is coupled to the bath. At
low frequencies, the curvature of the CODS is always negative for a sharp
cutoff while it is positive for an algebraic cutoff in the regime where the
specific heat becomes negative.

Quantitative differences appear in the details of the relevant quantities, e.g.\
the expressions for the mass renormalization or the critical damping strength.
While the general form of the expressions for the two types of cutoff are
similar, important differences are due to the fact that for a given algebraic
cutoff, the allowed values of the low-frequency exponent $s$ are limited by the
conditions (\ref{eq:range_s_p}) or (\ref{eq:range_s_p_restricted}). As a
consequence, the mass renormalization and the amplitude of the high-temperature
quantum corrections for the algebraic cutoff are nonmonotonic and divergent at
$s=2p$. In contrast, the sharp cutoff is sufficiently strong to allow for
arbitrary values of $s$.  The corresponding mass renormalization and the
quantum corrections of order $1/T^2$ vanish in the limit of large exponents
$s$.

Finally, we have seen in the discussion of the leading low-temperature corrections
that while the critical damping strength typically determines whether the specific heat
can fall below its zero-temperature value, this is not always the case. For the
sharp cutoff, the results shown in Fig.~\ref{fig:c_lowt_1}(b) demonstrate that
a negative specific heat in the Ohmic case can appear even before the critical
damping strength is reached.

\acknowledgments
The authors would like to thank Peter H{\"a}nggi and Peter Talkner for
stimulating discussions. One of us (U.W.) has received financial support from
the Deutsche Forschungsgemeinschaft through SFB/TRR~21.  GLI is grateful to the
Laboratoire Kastler Brossel in Paris for its hospitality during the preparation
of the manuscript.


\begin{thebibliography}{00}

\bibitem{Haenggi08}
 P. H{\"a}nggi, G.-L. Ingold, and P. Talkner,
 New J. Phys. \textbf{10}, 115008 (2008).

\bibitem{Florens04}
 S. Florens and A. Rosch,
 Phys. Rev. Lett. \textbf{92}, 216601 (2004)

\bibitem{Ingold09b}
 G.-L. Ingold, A. Lambrecht, and S. Reynaud,
 Phys. Rev. E \textbf{80}, 041113 (2009).

\bibitem{Zitko09}
 R. {\v Z}itko and T. Pruschke,
 Phys. Rev. B \textbf{79}, 012507 (2009).

\bibitem{Campisi09}
 M. Campisi, P. Talkner, and P. H{\"a}nggi,
 J. Phys. A: Math. Theor. \textbf{42}, 392002 (2009)

\bibitem{Campisi10}
 M. Campisi, D. Zueco, and P. Talkner,
 Chem. Phys. \textbf{375}, 187 (2010)

\bibitem{Sulaiman10}
 A. Sulaiman, F. P. Zen, H. Alatas, and L. T. Handoko,
 Phys. Rev. E \textbf{81}, 061907 (2010)

\bibitem{Merker12}
 L. Merker and  T. Costi,
 Phys. Rev. B \textbf{86}, 075150 (2012).

\bibitem{Haenggi06}
 P. H{\"a}nggi and G.-L. Ingold,
 Acta Phys. Pol. B \textbf{37}, 1537 (2006).

\bibitem{Ingold09}
 G.-L. Ingold, P. H{\"a}nggi, and P. Talkner,
 Phys. Rev. E \textbf{79}, 061105 (2009).

\bibitem{Ingold12}
 G.-L. Ingold,
 Eur. Phys. J. B \textbf{85}, 30 (2012).

\bibitem{Weiss12}
 U. Weiss,
 \textit{Quantum Dissipative Systems}, 4$^\text{th}$ edition
 (World Scientific, Singapore) 2012.

\bibitem{Hakim85}
 V. Hakim and V. Ambegaokar,
 Phys. Rev. A \textbf{32}, 423 (1985).

\bibitem{Grabert88}
 H. Grabert, P. Schramm, and G.-L. Ingold,
 Phys. Rep. \textbf{168}, 115 (1988).

\bibitem{NISTHandbook10}
 F. W. J. Olver, D. W. Lozier, R. F.  Boisvert,
 and C.W. Clark (eds.),
 \textit{NIST Handbook of Mathematical Functions}
 (Cambridge University Press, New York), 2010.

\bibitem{Schramm87}
 P. Schramm and H. Grabert,
 J. Stat. Phys. \textbf{49}, 767 (1987).

\bibitem{Bao05}
 J.-D. Bao, P. H{\"a}nggi, and Y.-Z. Zhuo,
 Phys. Rev. E \textbf{72}, 061107 (2005).

\bibitem{Grabert87}
 H. Grabert, P. Schramm, and G.-L. Ingold,
 Phys. Rev. Lett. \textbf{58}, 1285 (1987).

\bibitem{Spreng13}
 B. Spreng, G.-L. Ingold, and U. Weiss,
 EPL \textbf{103}, 60007 (2013).

\bibitem{Adamietz14}
 R. Adamietz, G.-L. Ingold, and U. Weiss,
 to appear in Eur. Phys. J. B.

\bibitem{Ford85}
 G. W. Ford, J. T. Lewis, and R. F. O'Connell,
 Phys. Rev. Lett. \textbf{55}, 2273 (1985)

\bibitem{Ford88}
 G. W. Ford, J. T. Lewis, and R. F. O'Connell,
 Ann. Phys. (N.Y.) \textbf{185}, 270 (1988)
\end{thebibliography}
\end{document}